\documentclass[aps,pra,amsmath,amssymb,floatfix,twocolumn, amsmath, superscriptaddress, twocolumn]{revtex4}
\usepackage{multirow}
\usepackage{bbold}
\usepackage{subfigure}
\usepackage{color}
\usepackage{mathrsfs}
\usepackage{hyperref}
\usepackage{comment}

\newcommand{\bvec}[1]{{\mathbf #1}}

\usepackage{amsfonts, relsize, color}
\usepackage{graphics}
\usepackage{graphicx}
\usepackage{subfigure}
\usepackage{hyperref}
\usepackage{color}

\begin{document}
\title{From Birefringent Electrons to a Marginal or Non-Fermi Liquid of Relativistic Spin-1/2 Fermions: An Emergent Superuniversality}

\author{Bitan Roy}
\affiliation{Max-Planck-Institut f$\ddot{\mbox{u}}$r Physik komplexer Systeme, N$\ddot{\mbox{o}}$thnitzer Stra. 38, 01187 Dresden, Germany}

\author{Malcolm P. Kennett}
\affiliation{Department of Physics, Simon Fraser University, 8888 University Drive, Burnaby, British Columbia, V5A 1S6, Canada}

\author{Kun Yang}
\affiliation{National High Magnetic Field Laboratory and Department of Physics, Florida State University, Tallahassee, Florida 32306, USA}

\author{Vladimir Juri\v ci\' c}
\affiliation{Nordita, KTH Royal Institute of Technology and Stockholm University, Roslagstullsbacken 23,  10691 Stockholm,  Sweden}

\date{\today}

\begin{abstract}
We present the quantum critical theory of an interacting nodal Fermi-liquid of quasi-relativisitc pseudo-spin-3/2 fermions that have a
non-interacting \emph{birefringent} spectrum with \emph{two} distinct Fermi velocities. When such quasiparticles interact with gapless bosonic degrees of freedom that mediate either the long-range Coulomb interaction or its short range component (responsible for spontaneous symmetry breaking), in the deep infrared or quantum critical regime in two dimensions the system is respectively described by a \emph{marginal-} or a \emph{non-Fermi liquid} of relativistic spin-1/2 fermions (possessing a \emph{unique} velocity), and is always a \emph{marginal Fermi liquid} in three dimensions. We consider a possible generalization of these scenarios to fermions with an arbitrary half-odd-integer spin, and conjecture that critical spin-1/2 excitations represent a \emph{superuniversal} description of the entire family of interacting quasi-relativistic fermions.
\end{abstract}

\maketitle

\emph{Introduction}:
All fermions in the Standard Model have  spin $1/2$, however higher spin particles, such as the \emph{gravitino}, a charge-neutral spin-3/2 fermion, have been postulated in theories such as supergravity \cite{Weinberg}. An important recent advance in condensed matter physics is the discovery of (quasi)-relativistic spin-1/2 fermions in graphene~\cite{graphene-review}, on the surface of topological insulators~\cite{Hasan-Kane-review, Qi-Zhang-review, Bansil-review}, in Weyl materials~\cite{Armitage-review} and in topological superconductors~\cite{Elliott-review}. It is also conceivable to realize higher spin fermions as \emph{emergent quasiparticles} in various solid state systems in the vicinity of band-touching points~\cite{luttinger, kennett-1, kennett-2, kennett-3, Dora, Watanabe, Lan-1, Lan-2, Bradlyn, liangfu, Ezawa, Chen, guo-numerics}, which can be either symmetry protected or correspond to a fixed point description of a quantum phase transition between two topologically distinct insulators.

Pseudo-spin-3/2 fermions~\cite{footnote-pseudo} can be found in the close proximity of \emph{linear} or \emph{bi-quadratic} touching of valence and conduction bands~\cite{luttinger}. We focus on the former situation where the quasiparticles display  a \emph{birefringent} spectrum with two distinct Fermi velocities, and therefore manifestly break Lorentz symmetry. Such fermions can be realized from simple tight-binding models on a two-dimensional generalized $\pi$-flux square lattice \cite{kennett-1,kennett-2,kennett-3}, honeycomb lattices \cite{Dora, Watanabe}, shaken optical lattices~\cite{Lan-1, Lan-2}, as well as in three-dimensional strong spin-orbit coupled systems~\cite{Bradlyn,Ezawa}, such as anti-perovskites~\cite{liangfu} and the CaAgBi family of materials~\cite{Chen}. In the present Letter we venture into the largely unexplored territory \cite{kennett-3,Dora,guo-numerics} that encompasses the response of such peculiar gapless fermionic excitations and their stability in the presence of electronic interactions.

We now provide a brief summary of our main findings.
Irrespective of their materials origin and dimensionality of the system, we show that the optical conductivity of non-interacting spin-3/2 fermions at zero temperature is \emph{identical} to that of pseudo-relativistic spin-1/2 fermions.
When spin-3/2 fermions interact with massless bosonic degrees of freedom, which mediate either the long-range Coulomb interaction or its short-range component, in the deep infrared or quantum critical regime,  a \emph{marginal Fermi-liquid} of effective spin-1/2 fermions emerges, featuring logarithmic corrections to its Fermi velocity in three dimensions.
By contrast, in two spatial dimensions, the system respectively hosts a marginal Fermi liquid or a \emph{non-Fermi liquid} of relativistic spin-1/2 fermions. At the non-Fermi liquid fixed point the residue of quasi-particle pole vanishes in a power-law fashion, and the ordered phase for a strong repulsive interaction represents a \emph{quadrupolar} charge- or spin-density-wave, while it is a spin-singlet $s$-wave paired state in the case of a strong attractive interaction. Finally, based on the form of the Hamiltonian for arbitrary half-odd-integer spin relativistic fermions, we conjecture that the corresponding nodal liquid is ultimately described in terms of \emph{critical spin-1/2 fermions}, which promotes these excitations as the \emph{superuniversal} description of the entire family of interacting quasi-relativistic fermions.


\emph{Hamiltonian}: The low-energy Hamiltonian describing a collection of quasi-relativistic pseudo-spin-3/2 fermions is given by ${\tilde\eta} \otimes H_{\frac{3}{2}} ({\bf k})$, where
\begin{equation}
H_{\frac{3}{2}} ({\bf k})= v \sum^{d}_{j=1} \bigg[ \Gamma_{j0} k_j + \beta \Gamma_{0j} k_j \bigg],
\label{Eq:32general}
\end{equation}
in dimensions $d=2$ and $d=3$ with $\tilde{\eta}$ defined below.
The four component Hermitian matrices are $\Gamma_{\mu \nu}=\sigma_\mu \otimes \sigma_{\nu}$, where
$\left\{ \sigma_\mu \right\}$, $\mu = 0, 1, 2, 3$ are the standard two dimensional Pauli matrices, and
 ${\bf k}=\left( k_1, \cdots, k_d \right)$. The \emph{isotropic} spectrum
 of $H_{\frac{3}{2}} ({\bf k})$ is given by $\pm v \left( 1 \pm \beta \right) |{\bf k}| = v_\pm |\bvec{k}|$, displaying
\emph{birefringence} with two effective Fermi velocities $v_\pm =v (1 \pm \beta)$, where $\beta$ is the birefringence
parameter.
Notice the spectra of $\tilde{H}_{3/2}({\bf k})=v \left( {\bf S} \cdot {\bf k} \right)$, namely $\pm v (3,1)|{\bf k}|/2$, where ${\bf S}$ are spin-3/2 matrices, are recovered for $\beta=1/2$.
For $\beta=0$ we recover two \emph{decoupled} copies of quasi-relativistic pseudo-spin-1/2 fermions,
similar to the situation in monolayer graphene~\cite{graphene-review} or the regular $\pi$-flux square lattice~\cite{marston-affleck}.
For $\beta=1$ the spectrum accommodates two flat bands and a Dirac cone, as found for the Lieb
lattice~\cite{Dagotto}. The Hamiltonian in Eq.~(\ref{Eq:32general}) permits the most general birefringent structure. Here, we restrict ourselves to $|\beta| <1$. In $d=2$ the
Hamiltonian Eq.~(\ref{Eq:32general}) describes low-energy excitations in a generalized $\pi$-flux square
lattice, with ${\tilde\eta} \equiv \eta_0$ \cite{kennett-1,kennett-2,kennett-3},
while Pauli matrices $\left\{ \eta_\mu \right\}$ act on the spin indices.
In contrast, in three dimensions $H_{\frac{3}{2}} ({\bf k})$ could describe spin-3/2 Weyl excitations
in a system with strong spin-orbit coupling, with ${\tilde\eta} \equiv \eta_3$, and the
$\left\{ \eta_\mu \right\}$ operating on valley indices. Independent of these
microscopic details the minimal representation of quasi-relativistic spin-3/2 fermions
is \emph{four dimensional}, in contrast to spin-1/2 fermions for which the
minimal representation is \emph{two dimensional}.


\emph{Optical conductivity}: We first focus on the \emph{response}  of spin-3/2 fermions to an external
electro-magnetic field and compute the optical conductivity at temperature $T=0$. The current operator
in the $l^{th}$ direction is $j_l= v \left( \Gamma_{0l} + \beta \Gamma_{l0}\right)$, where $l=1, \cdots, d$.
To extract the optical conductivity in a $d$-dimensional non-interacting system we compute the polarization bubble~\cite{SM}
\begin{equation}
\Pi^{(d)}(i\Omega) = -\frac{e^2_0}{\hbar d} \sum^{d}_{l=1} \bigg[ \Pi_{ll} (i \Omega) - \Pi_{ll} (i \Omega=0) \bigg],
\end{equation}
where $e_0$ is the external electronic test charge and
\begin{eqnarray}
\Pi_{ll} (i \Omega) ={\rm Tr} \int \frac{d^d{\bf k}}{(2 \pi)^d} \int^{\infty}_{-\infty} \frac{d\omega}{2 \pi} j_{l} G_{\bf k}\left( i \omega +i \Omega \right) j_l G_{\bf k} \left( i \omega \right).
\end{eqnarray}
Here $G_{\bf k}(i\omega)$ is the fermionic Greens function in terms of Matsubara frequency ($i\omega$). We then perform analytic continuation to real frequencies ($\omega$) via $i \Omega \to \omega+ i \delta$ and use the Kubo formula to obtain the optical conductivity
\begin{equation}
\sigma^{(2)}_{ll}(\omega)=\frac{e^2_0}{h} \frac{\pi N_f}{4}, \:\:\:
\sigma^{(3)}_{ll}(\omega)= \frac{e^2_0}{h} \frac{N_f \omega}{6 v},
\end{equation}
respectively in $d=2$ and $3$, where $N_f$ is the number of four-component spin-3/2 fermion species. Note
that the optical conductivity in a nodal Fermi liquid of spin-3/2 fermions with $N_f$ flavors is
\emph{identical} to that of $2 N_f$ component spin-1/2 quasi-relativistic Dirac or Weyl fermions~\cite{graphene-review, Armitage-review},
since $\sigma^{(d)}_{ll}(\omega)$ does not depend on $\beta$. Therefore, the above example already
indicates that  the birefringence parameter ($\beta$) may not be important for the physical properties
of this state at $T=0$, in the absence of an infrared cut-off. We now include the interactions to study their effects on spin-3/2 fermions, and show that their main role is to make the birefringence parameter irrelevant and restore Lorentz symmetry.


\begin{figure}[t!]
\subfigure[]{
\includegraphics[width=4.0cm]{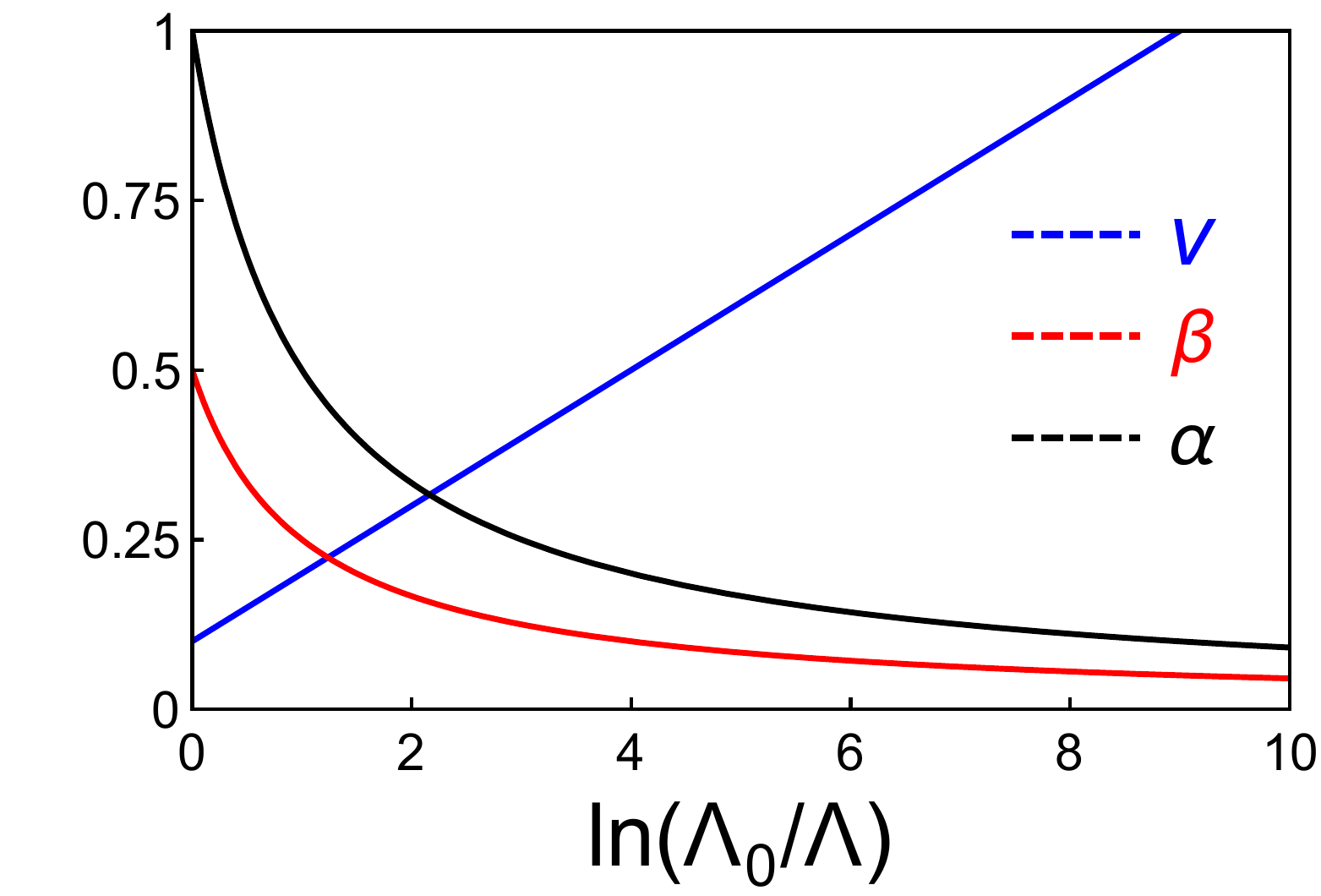}
\label{RGFlow:Coulomb2}}
\subfigure[]{
\includegraphics[width=4.0cm]{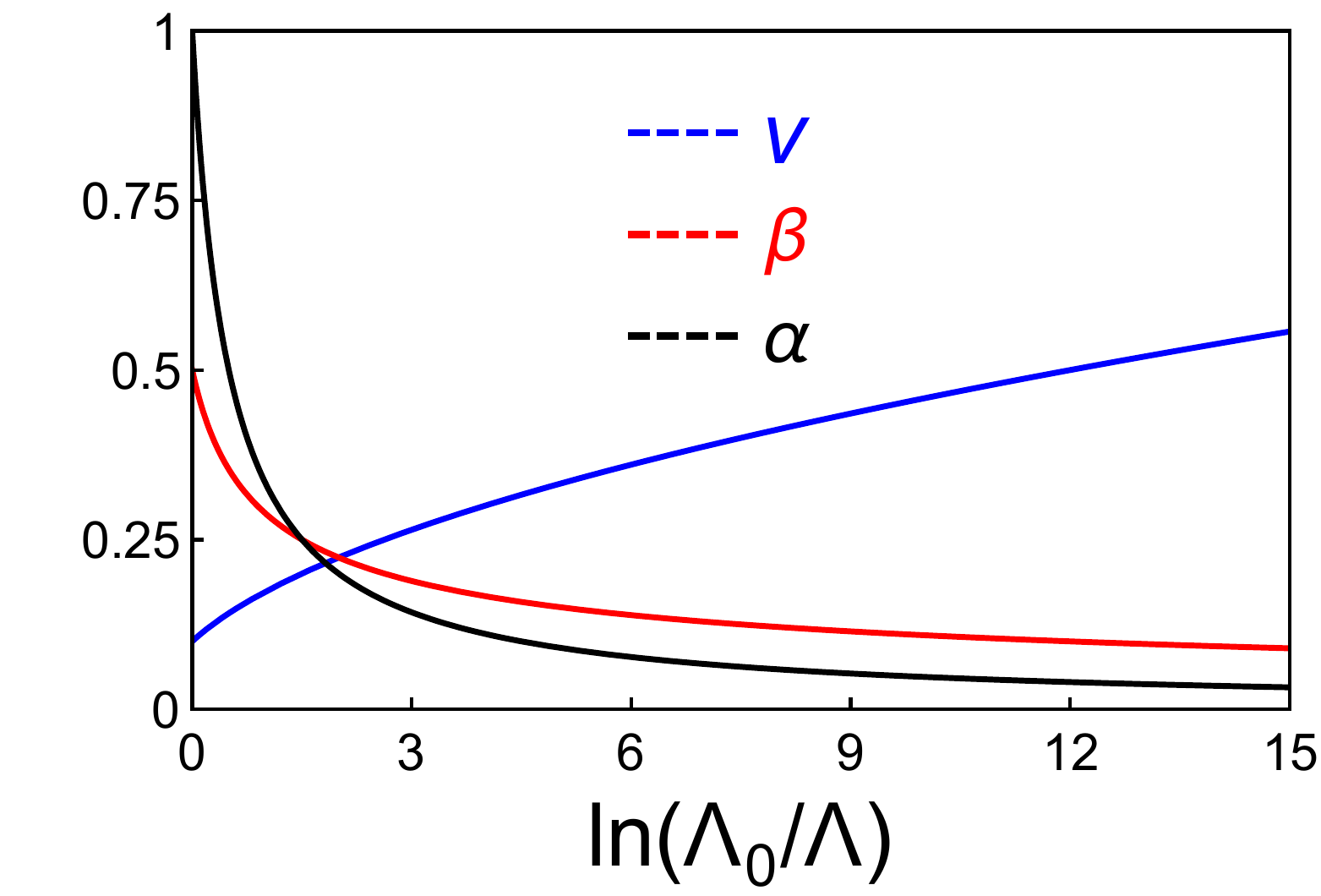}
\label{RGFlow:Coulomb3}
}
\caption{Renormalization group (RG) flow of the mean Fermi velocity ($v$), birefringence parameter ($\beta$)
and fine structure constant ($\alpha$) for quasi-relativistic pseudo-spin-3/2 fermions in the presence
of an instantaneous long-range Coulomb interaction in (a) $d=2$ and (b) $d=3$. The bare values of the
parameters are chosen to be $v_0=0.1$, $\beta_0=0.5$, $\alpha_0=1$.
}
\end{figure}

\emph{Long-range Coulomb interaction}: We focus first on the long-range tail of the density-density
Coulomb interaction, and neglect the retarded current-current interaction, since in a
condensed matter system $v_\pm \ll c$ (the speed of light).
The imaginary-time ($\tau$) action capturing the instantaneous Coulomb interaction is
\begin{equation}
S_{C}=\int d\tau d^d{\bf r} \; d^d{\bf r}^\prime \rho(\tau,{\bf r}) V({\bf r}-{\bf r}^\prime) \rho(\tau,{\bf r}^\prime),
\end{equation}
where $V({\bf r}-{\bf r}^\prime)=e^2/|{\bf r}-{\bf r}^\prime|$ and $\rho(\tau,{\bf r})$
is the fermionic density. In reciprocal space the Coulomb interaction $V({\bf k}) \sim e^2/|{\bf k}|^{d-1}$
is an analytic (a non-analytic) function of momentum in three (two) dimensions.
Therefore, only in $d=3$ is charge dynamically screened by fermions,
since the fermion bubble can only yield corrections that are analytic in momentum~\cite{pallab, hosur, sankar, juricic-roy, vozmediano, isobe-nagaosa, roy-juricic-herbut}.
From the leading order fermionic and bosonic (soft gauge field mediated by Coulomb interaction)
self-energy corrections, we arrive at the following renormalization group flow equations~\cite{SM}
\begin{equation}~\label{Eq:CoulombFlow}
\frac{dv}{d \ell}= \frac{\alpha v}{C_d} \equiv \frac{e^2}{C_d}, \:
\frac{d\beta}{d \ell}=- \frac{\alpha \beta}{C_d}, \:
\frac{d\alpha}{d \ell}=-(1+ N_f \delta_{d,3}) \; \frac{\alpha^2}{C_d},
\end{equation}
where $\alpha=e^2/v$ is the fine structure constant, $C_2=8 \pi$, $C_3=6 \pi^2$, and $\ell \equiv \ln (\Lambda_0/\Lambda) > 1$
is the logarithm of the running renormalization group scale, with $\Lambda_0$ the ultraviolet cut-off, while $\Lambda$ is the
running scale. To leading order, the birefringent part of $H_{\frac{3}{2}}({\bf k})$ remains \emph{marginal} in the presence
of Coulomb interactions, namely $d(\beta v)/d\ell=0$, while the mean Fermi velocity ($v$) receives a $\beta-$independent
logarithmic correction making $v$ a marginally relevant parameter.

In a two or three dimensional interacting quasi-relativistic liquid of spin-3/2 fermions the mean Fermi velocity ($v$)
thus increases \emph{logarithmically}, while the birefringent parameter ($\beta$) decreases, also logarithmically,
but the parameter $\beta v$ remains \emph{marginal}. Ultimately, in the deep infrared regime we recover \emph{effectively}
two decoupled copies of relativistic spin-1/2 fermions, since $v_+-v_- \ll v$ yielding $\beta v \ll v$.
We dub such a state an {\it effective marginal Fermi-liquid} of spin-$1/2$ fermions.
As a result, the fine structure constant ($\alpha$) also decreases logarithmically slowly.
These scenarios for $d=2$ and $d=3$ respectively are shown in Figs.~\ref{RGFlow:Coulomb2} and \ref{RGFlow:Coulomb3}.

\begin{figure}[t!]
\subfigure[]{
\includegraphics[width=3.9cm]{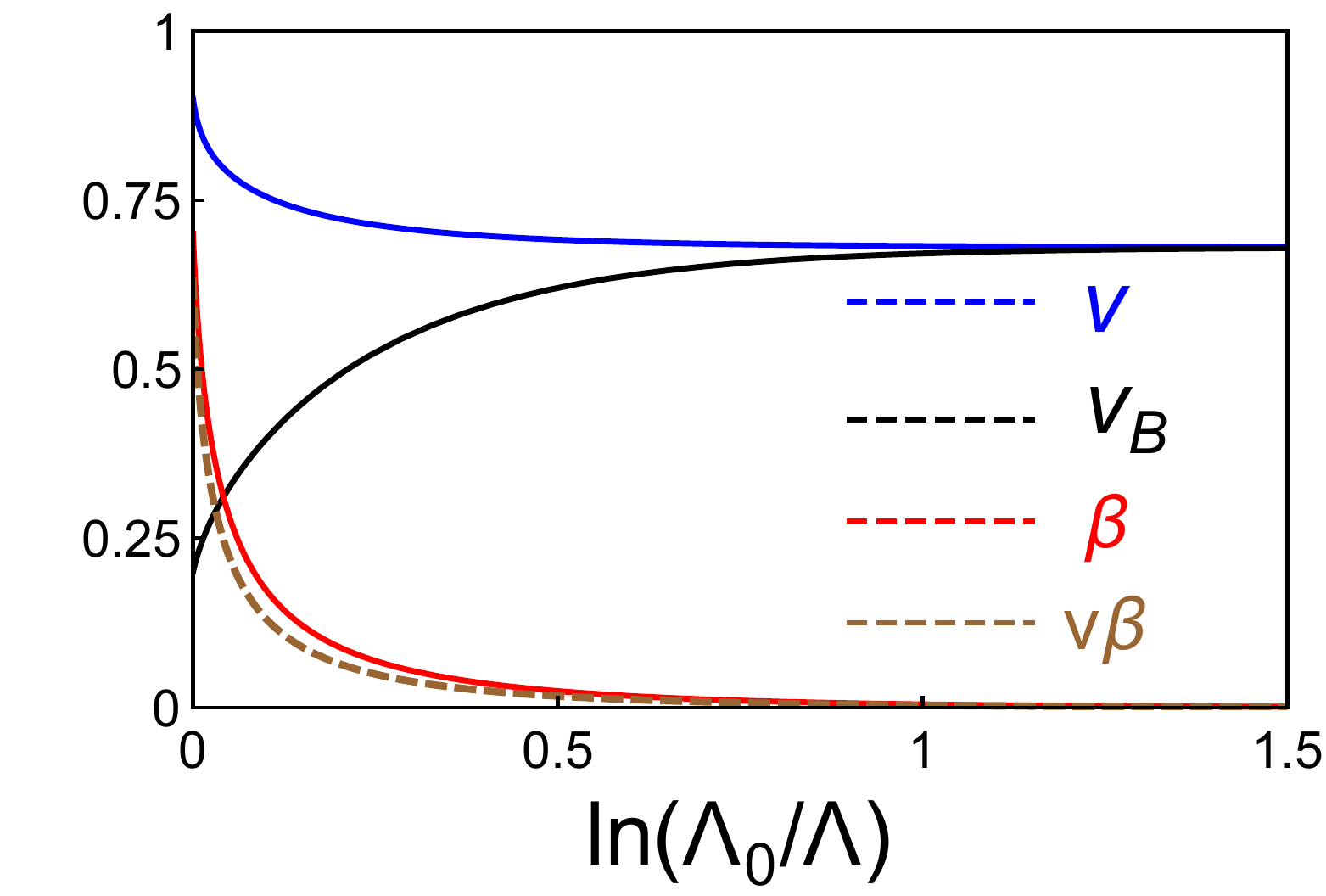}
~\label{yukawa1}
}
\subfigure[]{
\includegraphics[width=3.9cm]{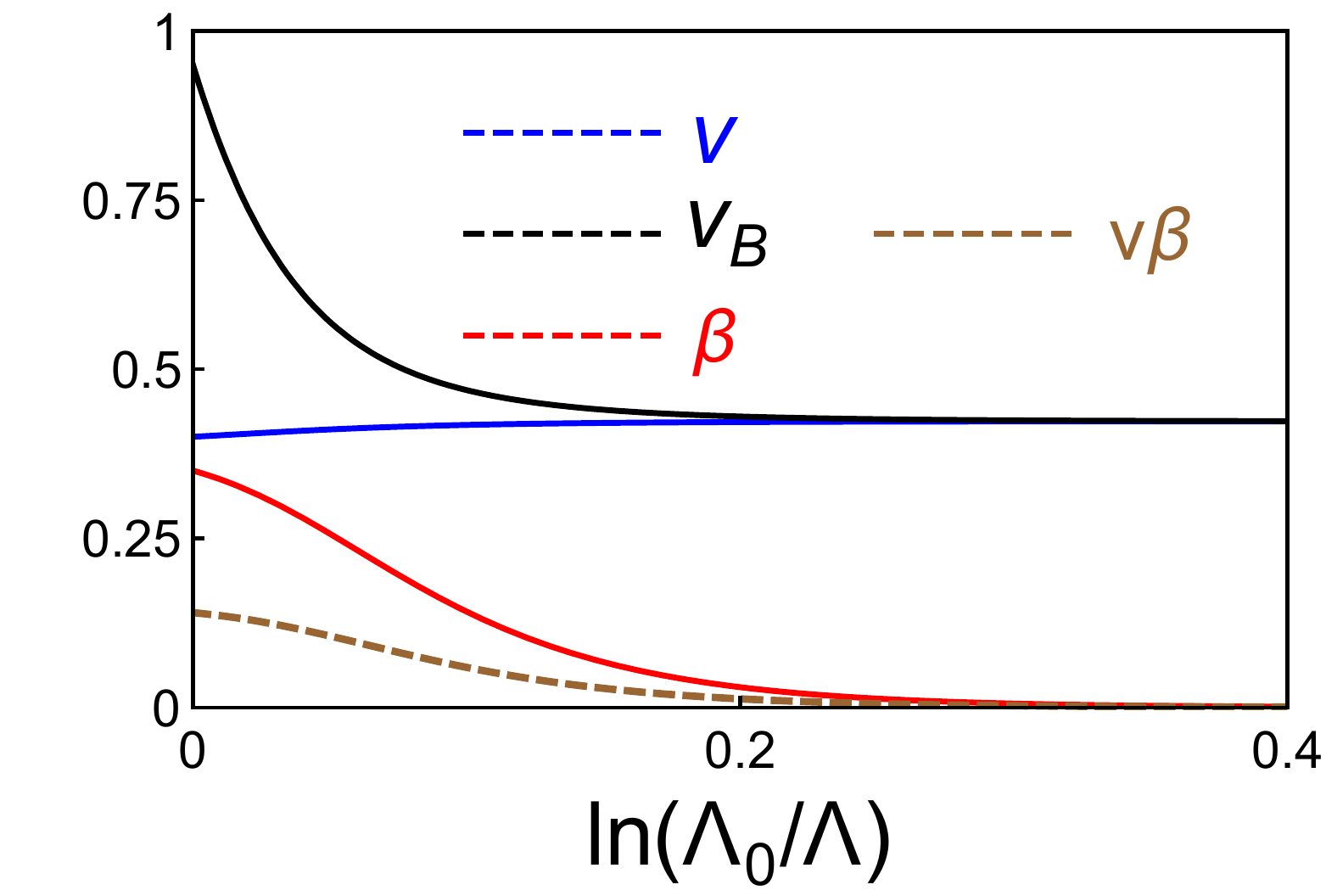}
~\label{yukawa2}
}
\caption{RG flow of two Fermi velocities ($v$ and $v \beta$), bosonic ($v_B$) velocity and the birefringence parameter ($\beta$) in the presence of
Yukawa interaction ($g^2 \sim \epsilon$) when (a) $v_0>v_{B,0}$ and (b) $v_{B,0}>v_{0}$, for $N_f=N_b=1$. For panel (a) we choose $v_0=0.9, v_{B,0}=0.2, \beta_0=0.7$, $v_0 \beta_0 =0.63$
and for (b) $v_0=0.4, v_{B,0}=0.95, \beta_0=0.35$, $v_0 \beta_0 =0.14$.
}~\label{Flow-numerical:Yukawa}
\end{figure}

Due to our neglect of the retarded part of the long range interaction, the enhancement of the mean Fermi velocity ($v$)
is \emph{unbounded}. However, as $v$ increases, the current-current interaction (originally suppressed by
$v_\pm/c \ll 1$) can no longer be neglected and ultimately the flow of the Fermi velocity stops at the speed of light ($c$),
leading to the restoration of genuine \emph{Lorentz symmetry}~\cite{vozmediano, isobe-nagaosa, roy-juricic-herbut}.
Although the parameter $\beta v$ is initially \emph{marginal},
based on the results with retarded short-range interaction mediated by a bosonic order-parameter field,
which we discuss below, we suspect that \emph{ultimately} $\beta v$ (originally a marginal or scale ($\ell$) invariant quantity in the presence of only instantaneous Coulomb interaction) becomes \emph{irrelevant}
 with the inclusion of retarded long-range current-current interaction, implying $\beta v \to 0$ as $\ell \to \infty$ (see brown curves in Fig.~\ref{Flow-numerical:Yukawa} for qualitative comparison). Therefore,
at the lowest energy scales, once the full electromagnetic interaction is accounted for,
only spin-$1/2$ fermions survive. However, the length scale ($\ell_\ast$) at which the current-current interaction
becomes important is extremely large ($\ell_\ast \gg 1$), and logarithmically slow growth of the Fermi velocity makes
it impossible to access such a regime in any real system. We leave this issue of definite fundamental importance but
pure academic interest for future investigation.
In $d=3$, besides the Lorentz symmetric fixed point it is also conceivable to find an $O_h$ (cubic) symmetric interacting fixed point in a crystalline environment. However, the Lorentz symmetric fixed point has a larger basin of attraction~\cite{isobe-fu}.


\emph{Proximity to a Mott transition}: Finally, we address the quantum critical description of a collection of
spin-3/2 fermions, residing near a Mott transition, driven into an insulating phase by the short range parts of the Coulomb interaction
(such as those appearing in an extended Hubbard model)~\cite{kennett-3,guo-numerics}. The density of states vanishes as $\varrho(E) \sim |E|$, hence
any ordering takes place at a finite coupling through a quantum phase transition. For spinless fermions in $d=2$
there exists one matrix, namely $\Gamma_{33}$, that \emph{anticommutes} with
$H_{\frac{3}{2}}({\bf k})$.  Thus, in an ordered phase with $\langle \Psi^\dagger \eta_0 \otimes \Gamma_{33} \Psi \rangle \neq 0$
the quasiparticle spectrum is fully and \emph{uniformly} gapped. Hence, at $T=0$ the propensity toward such an ordering is
energetically favored since it maximally lowers  the free-energy.
Such an ordered phase represents a \emph{quadrupolar charge-density-wave}, since $\Gamma_{33}=(2 S^2_z-S^2_x-S^2_y)/3$,
when $S_z={\rm diag}(3,1,-1,-3)/2$.
The inclusion of spin degrees of freedom leads to a competing order, also representing a massive ordered phase,
where $\langle \Psi^\dagger \vec{\eta} \otimes \Gamma_{33} \Psi \rangle \neq 0$ describes a \emph{quadrupolar spin-density-wave},
with $\vec{\eta}=\left( \eta_1,\eta_2,\eta_3\right)$. Both spin- and charge-density-wave phases break the discrete
particle-hole symmetry generated by $\Gamma_{33}$, as $\left\{H_{\frac{3}{2}}({\bf k}), \Gamma_{33} \right\}=0$,
while the former one also breaks the continuous $SU(2)$ spin-rotational symmetry generated by
$\vec{\eta}\otimes \Gamma_{00}$, hence the ordered phase is accompanied by \emph{two} Goldstone modes.
These two phases can be realized for sufficiently strong onsite~\cite{guo-numerics} and nearest-neighbor~\cite{kennett-3} repulsion respectively
in a $\pi$-flux square lattice. By contrast, in $d=3$ there is no matrix that anti-commutes with $H_{\frac{3}{2}}({\bf k})$
 and single-flavored spin-3/2 Weyl fermions cannot be gapped out. Nevertheless, if we account for valley degrees of
freedom then spin-3/2 Weyl fermions can be gapped out by spontaneously breaking  \emph{translational}
symmetry (generated by $\eta_3 \otimes \Gamma_{00}$), and the ordered phase is characterized by
$\langle \Psi^\dagger \vec{\eta}_\perp \otimes \Gamma_{00} \Psi \rangle \neq 0$, with
$\vec{\eta}_\perp=\left( \eta_1, \eta_2 \right)$. The same order parameter can
represent a spin-singlet $s$-wave pairing for spin-3/2 fermions in $d=2$ (with an appropriate redefinition of the
spinor basis). However, the fate of an emergent Lorentz symmetry close to a Mott transition for two or
three-dimensional linearly dispersing spin-3/2 fermions is insensitive to these details.

The imaginary time action describing such a quantum phase transition is given by ${\mathcal S}=\int d\tau d^d{\bf r}
\left( {\mathcal L}_f +{\mathcal L}_Y+{\mathcal L}_b \right)$, where ${\mathcal L}_f$ describes the dynamics of
gapless spin-3/2 fermionic excitations (represented by a spinor field $\Psi$) with
\begin{equation}
{\mathcal L}_f=\Psi^\dagger \left[ \partial_\tau + {\tilde\eta} \otimes H_{\frac{3}{2}} ({\bf k} \to -i \nabla) \right] \Psi,
\end{equation}
where ${\tilde\eta}=\eta_0$ or $\eta_3$. The coupling between gapless bosonic and fermionic degrees of freedom is captured by
\begin{equation}
{\mathcal L}_{Y}= g \sum^{N_b}_{\alpha=1} \Phi_\alpha \Psi^\dagger M_\alpha \Psi,
\end{equation}
where $g$ is the Yukawa coupling, and $N_b$ counts the number of real components of the bosonic order-parameter field.
Therefore, $N_b=1(3)$ for a quadrupolar charge(spin)-density-wave in $d=2$, and
$N_b=2$ for a translational symmetry breaking charge-density-wave in $d=3$ and $s$-wave pairing in $d=2$.
The Hermitian matrix $M_\alpha$ always anti-commutes with the non-interacting Hamiltonian $\tilde{\eta} \otimes H_{\frac{3}{2}}({\bf k})$.
The dynamics of the order-parameter bosonic field is captured by an appropriate relativistic-like $\Phi^4$ theory
\begin{equation}
L_b=\sum_{\alpha=1}^{N_b} \left[ -\frac{1}{2}\Phi_\alpha \left(\partial^2_\tau +v^2_B\sum^{d}_{j=1}\partial_\mu^2-m^2_b \right) \Phi_\alpha + \frac{\lambda}{4!} [\Phi_\alpha^2]^2 \right],
\end{equation}
where $\lambda$ is the four-boson coupling, $m^2_b$ (the bosonic mass) is the tuning parameter
for the transition with $m^2_b=0$ at the quantum critical point, and $v_B$ is the characteristic velocity of the
bosonic field. Note that both Yukawa and four-boson couplings are \emph{dimensionless} in $d=3$.
Therefore, the critical behavior of the above field theory can be addressed by performing a
controlled $\epsilon$-expansion about three spatial dimensions, with $\epsilon=3-d$~\cite{zinn-justin}.
From the leading order self-energy corrections we arrive at the following flow equations for the
two velocities ($v$ and $v_B$) and the birefringence parameter ($\beta$)~\cite{SM}
\begin{widetext}
\begin{equation}~\label{Eq:YukawaFlow}
\frac{d v}{d \ell} =- 2 N_b v g^2 A(v,v_B, \beta), \quad
\frac{d \beta}{d \ell}=-2 N_b \beta g^2 S(v,v_B, \beta), \quad
\frac{d v_B}{d \ell} = \frac{N_f g^2 v_B}{2 v^3 (1-\beta^2)} \:
\left[ C(v,v_B, \beta)- \frac{1}{1-\beta^2} \right],
\end{equation}
with $X(v,v_B,\beta) \equiv X$ for $X=A,S,C$ and where
\begin{equation}
A=\frac{2 \left( v-v_B\right) \left( v+v_B\right)^2 + 4 v^3 \beta^2}{3 v v_B \left[ \left( v+v_B\right)^2 -v^2 \beta^2\right]}, \:\:
S=\frac{2 \left[ 4 v v_B \left( v+v_B\right) +v^3_B + v^3(1-\beta^2) \right]}{3 v v_B \left[ \left( v+v_B\right)^2 -v^2 \beta^2\right]}, \:\:
C=\frac{v^2}{v^2_B} -\beta^2 \left[\frac{4}{(1-\beta^2)^2} -\frac{2 v^2}{3 v^2_B}\right].
\end{equation}
\end{widetext}
Even though it is a daunting task to solve the above coupled flow equations exactly,
valuable insights can be gained from their numerical solutions, displayed in Fig.~\ref{Flow-numerical:Yukawa}.

We note that regardless of whether $v>v_B$ [Fig.~\ref{yukawa1}] or $v_B>v$ [Fig.~\ref{yukawa2}] in the bare theory,
the Fermi and bosonic velocities approach each other, while the birefringent parameter $\beta$ flows to zero in the deep infrared
regime, but in this case also the \emph{parameter $\beta v$ separately flows to zero}. Hence, as the system approaches the boson-fermion coupled Yukawa fixed point,
it is effectively described by two \emph{decoupled} copies of spin-1/2 fermions and the Lorentz symmetry gets
restored since $v_+$ and $v_-$ approach a common velocity, $v,v_B \rightarrow {\tilde v}$, and $v \beta \to 0$ (see brown dashed line in Fig.~\ref{Flow-numerical:Yukawa}) as $\ell \to \infty$.
The coupled flow equations for the remaining two couplings in the $\beta=0$ and $m^2_b=0$ hyperplane take the form
\begin{eqnarray}~\label{RG:zippedgeneral}
\frac{dg^2}{d \ell} = \epsilon g^2- a_1 g^4,
\frac{d\lambda}{d\ell} = \epsilon \lambda -4 N_f g^2 \left[\lambda-6 g^2 \right] - \frac{a_3 \lambda^2}{6}, \nonumber  \\
\end{eqnarray}
and support only one quantum critical point located at
\begin{equation}~\label{FP:locationgeneral}
\left( g^2_\ast, \lambda_\ast \right)= \left( 1, \frac{3}{a_3} \left[ a_2+ \sqrt{a_2^2+ 16 N_f a_3}\right]\right) \frac{\epsilon}{a_1},
\end{equation}
where $a_1=2 N_f+4-N_b$, $a_2=4-2N_f-N_b$ and $a_3=N_b+8$. At this critical point both
fermionic and bosonic excitations possess non-trivial anomalous dimensions,
given by $\eta_f=N_b g^2_\ast/2$ and $\eta_b= 2 N_f g^2_\ast$ respectively,
responsible for the absence of sharp quasiparticles in its vicinity. Specifically,
the residue of the fermionic quasiparticle pole vanishes as $Z_\Psi \sim (m_f)^{\eta_\Psi/2}$,
where $m_f$ is the fermionic mass that vanishes  following a universal ratio $(m_b/m_f)^2 \sim \lambda_\ast/g_\ast$
as the critical point is approached from the ordered side. Therefore, in $d=2$ or $\epsilon=1$, $Z_\Psi$
vanishes in a \emph{power-law} fashion, and the Yukawa critical point accommodates
a \emph{relativistic non-Fermi liquid}, where gapless spin-1/2
fermionic and bosonic order-parameter excitations are strongly coupled.
By contrast, in $d=3$ or $\epsilon=0$, the critical phenomena at the transition are controlled by a Gaussian fixed point, located at $g^2_\ast=\lambda_\ast=0$, which exhibit mean-field behavior with logarithmic corrections due to the fact that the field theory is at its upper critical dimension. Consequently, the residue of quasiparticle pole vanishes only
\emph{logarithmically}, and the Yukawa fixed point hosts a \emph{relativistic marginal Fermi liquid}.


\emph{Discussion}: We demonstrate that when quasi-relativistic spin-3/2 fermions interact with bosonic degrees of freedom which  represent either a soft gauge field mediating the long-range or an order-parameter field mediating a short-range Coulomb interaction, in the deep infrared regime the system is described by either an effective marginal- or non-Fermi liquid of relativistic spin-1/2 excitations~\cite{footnote-1}.
One can envision  generalizing our analysis to spin-$s/2$ quasi-relativistic fermions with arbitrary odd value of $s$.
For spin-$s/2$ Weyl fermions the low-energy Hamiltonian can be written as $\tilde{H}_{\frac{s}{2}}({\bf k})=v \left( {\bf S} \cdot {\bf k} \right)$, where ${\bf S}$
are now spin-$s$/2 matrices, and the quasiparticle spectrum has $(s+1)/2$ effective Fermi velocities, given by $v_s=\left(1/2,3/2, \cdots, s/2 \right)v$. Even though we expect our conclusions to hold for any value of $s$, specifically when $s+1=2^N$, with $N$ an integer, a more general
\emph{multirefrigence} can be accommodated by the Hamiltonian
\begin{equation}
H_{\frac{s}{2}}({\bf k}) = v \sum^{d}_{j=1} k_j \bigg[ \Gamma_{j 0 \cdots 0} + \cdots + \beta_{N-1} \Gamma_{0 \cdots 0j} \bigg],
\end{equation}
similar to Eq.~(\ref{Eq:32general}). Here $\Gamma_{j 0 \cdots 0}=\sigma_j \otimes \sigma_0 \otimes \cdots \otimes \sigma_0$
and so on are $(s+1)$-dimensional Hermitian matrices. Note that for $d=2$ there exists a matrix, namely $\Gamma_{33\cdots 3}=\sigma_3 \otimes  \cdots \otimes \sigma_3$, that fully anti-commutes with $H_{\frac{s}{2}}({\bf k})$.
Therefore, we believe that our proposed critical descriptions for linearly dispersing spin-3/2 fermions are also
applicable for spin-$s/2$ fermions~\cite{footnote-2}. This would imply that critical relativistic spin-1/2 fermions (describing a marginal or non-Fermi liquid)
stand as an extremely sparse example of an \emph{emergent superuniversal} description of the entire family of interacting
quasi-relativistic spin-$s/2$ fermions in two and three dimensions. This conjectured property of relativistic fermions with
a half-odd-integer spin could be demonstrated numerically~\cite{guo-numerics} and possibly in experiments.
Finally, we note that the present discussion might have relevance to the observation that all
fermions in the Standard Model are described in terms of the spin-1/2 representation, and this feature could be
envisioned as an example of an \emph{emergent phenomenon}, analogous to the restoration of Lorentz symmetry~\cite{nielsen}.

\emph{Acknowledgments}: M. K. was supported by NSERC of Canada. K. Y. is supported by National Science Foundation Grants No. DMR-1644779 and No. DMR-1442366. B. R. is thankful to Nordita for hospitality.




\begin{thebibliography}{}
\bibitem{footnote-2} When the long-range tail of the Coulomb interaction dominates over its short-ranged pieces, a spin-$s/2$ system flows toward a marginal-Fermi liquid fixed point of spin-1/2 fermions. This outcome does not rely on the existence of the anti-commuting or mass matrix $\Gamma_{33\cdots 3}$.
\end{thebibliography}
\end{document}